\documentstyle[12pt,aasms4]{article}
\tighten
\begin{document}
\title{Integral Field Spectroscopy of Mrk 273: Mapping 10$^3$ 
       {\lowercase{km s}}$^{-1}$ 
       Gas Flows and an Off-Nucleus Seyfert 2 Nebula$^1$}

\vspace{0.5in}

\author{Luis Colina\altaffilmark{2}, Santiago Arribas\altaffilmark{3,4} \& 
Kirk D. Borne\altaffilmark{5}}

\affil{(2) \em Instituto de F\'{\i}sica de Cantabria (CSIC-UNICAN), Facultad de Ciencias,
39005 Santander, Spain (colina@ifca.unican.es)}

\affil{(3) \em Instituto de Astrof\'{\i}sica de Canarias, 38200 La Laguna, Tenerife
Spain (sam@ll.iac.es)}

\affil{(4) \em Consejo Superior de Investigaciones Cient\'{\i}ficas (CSIC)}

\affil{(5) \em Raytheon Information Technology and Scientific Services,  NASA Goddard Space
          Flight Center, Greenbelt, MD 20771, USA (borne@rings.gsfc.nasa.gov)}

\scriptsize
 
\vspace{1in}    
\footnote{
Based on observations with WHT operated on the island of La Palma by the ING in the
Spanish Observatorio del Roque de los Muchachos of the Instituto de Astrof\'{\i}sica 
de Canarias.}

\normalsize
 
\begin{abstract}

Integral field optical spectroscopy with the INTEGRAL fiber-based system
is used to map the extended ionized 
regions and 
gas flows in Mrk 273, one of the closest Ultraluminous Infrared galaxies 
(ULIRGs).

The H$\beta$ and [OIII]5007\AA\ maps show the presence of two distinct
regions separated by 4$''$ (3.1 kpc) along position angle (PA) 240$^\circ$. 
The northeastern region coincides with 
the optical nucleus of the galaxy 
and shows the spectral characteristics of LINERs. The southwestern region
is dominated by
[OIII] emission and is classified as a Seyfert 2. Therefore, in the
optical, Mrk 273 is an ultraluminous infrared galaxy with a LINER nucleus and 
an extended off-nucleus Seyfert 2 nebula. 

The kinematics of the [OIII] ionized gas shows (i) 
the presence of highly disturbed gas in the regions around the LINER nucleus, 
(ii) a high-velocity gas flow with a peak-to-peak amplitude of 2.4 $\times$ 10$^3$ 
km s$^{-1}$, and (iii) quiescent gas in the outer regions (at 3~kpc). 

We hypothesize that the high-velocity flow is the starburst-driven superwind generated 
in an optically obscured nuclear starburst, and that the quiescent gas is directly ionized by a 
nuclear source, like the ionization cones typically seen in Seyfert galaxies.

\end{abstract}


\section{INTRODUCTION}

Ultraluminous infrared galaxies (ULIRGs), with bolometric luminosities (L$_{IR} 
\simeq$ L$_{bol} \geq 10^{12}$ L$_{\odot}$),
are the brightest galaxies in the local universe. 
ULIRGs show signs of strong interactions or mergers and have large amounts 
of gas and dust that significantly obscure the nuclear ionizing sources 
(see Sanders \& Mirabel
1996 for a review). Mid-infrared spectroscopy has shown that ULIRGs with 
optical H\thinspace II- and LINER-like spectra are dominated by the
energy output from nuclear starbursts (Lutz {\it{et al.}}~1998; Genzel {\it{et al.}}~1999; Lutz, 
Veilleux, \& Genzel 1999). However, the increased fraction of active galactic nuclei (AGN) 
among the brightest ULIRGs (L$_{IR} 
\geq 10^{12.3}$ L$_{\odot}$) is taken as evidence for the presence of a dust-enshrouded quasar 
powering these galaxies, at least at the brightest end of the luminosity distribution
(Veilleux, Kim, \& Sanders 1999). 

Two-dimensional kinematical studies of a few ULIRGs (Mihos \& Bohun 1998) show that the ionized gas
in these galaxies has a wide variety of features affected by several factors,
such as the interaction stage, the gas content, 
and the effects of massive starbursts.  
    
The study of compact galaxies like ULIRGs showing complex velocity fields and ionization 
substructures requires the use of two-dimensional spectroscopy (i.e. integral field) to
characterize and map simultaneously  
the warm ionized gas, the cold dust, and the stellar component.   

Mrk 273 is one of the closest ULIRGs 
(z= 0.03778; Downes, Solomon \& Radford 1993).  It is
characterized by a long nearly straight tidal tail, a LINER 
(Veilleux {\it{et al.}}~1995) or
Seyfert 2 type optical spectrum (Sanders {\it{et al.}}~1988; Veilleux, Kim, \& Sanders 1999), and
possibly the presence of a nuclear starburst (Genzel {\it{et al.}}~1998; Lutz, Veilleux, \& Genzel 1999). 
The nuclear region
of Mrk 273 is very dusty and mottled, showing a very strong dust lane and a system of 
radially extended filaments similar to those seen in M82 
(Borne {\it{et al.}}~1999). 
Mrk 273 has two near-infrared nuclei separated by $\sim$1$''$
(Majewski {\it{et al.}}~1993; 
Knapen {\it{et al.}}~1997; 
Scoville {\it{et al.}}~1999).

In this letter we highlight the new results obtained for Mrk 273 integral field 
optical spectroscopy. The observations and reduction procedures are briefly mentioned in
section $\S$2,
and the results are presented in section $\S3$. Throughout the paper a
Hubble constant of 70 km s$^{-1}$ Mpc$^{-1}$ is assumed. 

\section{OBSERVATIONS AND DATA REDUCTION}

Integral field spectroscopy of Mrk 273 was obtained with the INTEGRAL system
(Arribas {\it{et al.}}~1998), a fiber-based spectrograph mounted at the 
Naysmith No.~1 
platform of the 4.2m William Herschel Telescope. Although this system has several fiber
bundles, the one used for the present observations  
consisted of 219 fibers, each 0.9$''$ in diameter
and covering a field-of-view of 16.5$'' \times 12.3''$. 
The spectra were taken using a 600 l/mmm grating, covering 
the 5000-7900 \AA\ range, and with an effective resolution of 4.8 \AA.. 
The total integration time was 4500 seconds, split into three separate integrations
of 1500 seconds each. The seeing had a value 
of $\sim$1.0$''$.

The reduction consists of two main steps: (i) basic reduction of the 
spectra, and (ii) generation of continuum and emission-line
images from the reduced spectra (see Colina \& Arribas 1999 for details). 
The 
[OIII]5007\AA\
profiles obtained {\it simultaneously} with INTEGRAL for the central 
region of Mrk 273  are shown in Figure 1 as an example of the kind of reduced
data obtained after step (i) is finished.  These represent only a small 
60\AA\ subsample of the full 5000-7900 \AA\ spectral coverage obtained 
at each of the 219 fiber locations.

\section{RESULTS AND DISCUSSION}

\subsection{A LINER Nucleus with an Off-nucleus Seyfert 2 Nebula}

Mrk 273 has been classified as a LINER (Veilleux {\it{et al.}}~1995)
and a Seyfert 2 (Veilleux, Kim \& Sanders 1999). These authors attribute the
change in the classification as due to differences in the extraction apertures,
suggesting therefore the existence of extended ionized gas in this galaxy.

Our data show that the stellar light distribution 
and the warm line-emitting gas 
have a rather different morphology (see Figure 2). The stellar blue light distribution
has a main nucleus, with a secondary faint nucleus about 4$''$ south of it. The stellar component 
is distributed along the North-South direction showing in the northern part an
elongation that marks the beginning of the well known tidal tail of this galaxy 
(see bottom left panel in Figure 2). 

By contrast, the ionized gas is distributed into two well differentiated high surface 
brightness regions located along PA=240$^\circ$, and with their emission peaks separated by 
about 4$''$ ({\it{i.e.}}, $\sim$ 3.1 kpc). Two additional regions, at lower surface brightness, 
are also identified. The first of these is located at about 3.5$''$ east of the main
northeastern region while the second one appears as a tongue about 2$''$ south of the 
main southwestern region (see left and middle top panels in Figure 2).

The northeastern H$\beta$-emitting region coincides with the blue 
continuum nucleus of the galaxy, while the southwestern [OIII]-emitting region 
lies outside the main stellar body of the galaxy and is not associated with 
any distinct source seen in the blue continuum. 
As in Arp 220 (Scoville {\it{et al.}}~1998) and other
ULIRGs (Evans 1999), the observed optical structure of Mrk 273 is very much affected 
by the dust lanes clearly visible in {\it{HST}} images (Borne {\it{et al.}}~1999). 
Comparisons of {\it{HST}} $I$-band and near-infrared images (Colina, unpublished) 
show that the optical nucleus is offset 
from the two well known near-infrared nuclei (Majewski et al 1993; Knapen {\it{et al.}}~1997; 
Scoville {\it{et al.}}~1999)  by $\sim$ 0.6$''$ (0.47 kpc) and $\sim$ 1.5$''$ (1.18 kpc) along 
PA45 and PA43, respectively.
Therefore the optical nucleus and
the ionized regions detected here correspond to clouds of 
interstellar gas being ionized by the obscured ionizing nuclear source(s) associated most likely
with the near-infrared nuclei. 

The two main line-emitting regions are characterized by different
excitation conditions as suggested by the observed variations in
their relative H$\beta$ and [OIII] surface 
brightnesses (Figure 2 \& 3).
Based on the observed H$\alpha$/H$\beta$ ratio, the northeastern
line-emitting region
has an internal extinction of 3.4 magnitudes in the visual.
After correcting for extinction, 
this region presents the spectral characteristics of a LINER
([OIII]/H$\beta$= 0.26 dex, [OI]/H$\alpha$= $-$0.84, [NII]/H$\alpha$= $-$0.09  
and [SII]/H$\alpha$= $-$0.33).
On the other hand, the high-excitation off-nuclear [OIII]-emitting region shows 
no evidence for internal extinction by dust and its
emission line ratios are characteristic of a Seyfert 2 
([OIII]/H$\beta$= 0.88 dex, [OI]/H$\alpha$= $-$0.96, [NII]/H$\alpha$= 
0.0 and [SII]/H$\alpha$= $-$0.32). 

Our new data show that, in the optical, Mrk 273 has a 
LINER nucleus and an off-nucleus Seyfert 2 extended nebula located at about 3.1 kpc 
southwest of the nucleus. However, the optical nucleus is offset from the two 
near-infrared nuclei. 
Moreover, {\it{HST}} NICMOS near-infrared images (Scoville {\it{et al.}}~1999) show that the northern 
near-infrared nucleus ({\it{i.e.}}, the one closest
to our LINER nucleus) is extended (starburst?) while the southern nucleus is a $K$-band
point-like source (AGN?), with a size less than 0.15$''$ ($\leq$ 120 pc).
Also, the extinction
derived in the mid-infrared (A$_V \geq$ 15; Genzel {\it{et al.}}~1998) is a factor
$\sim$5 larger than that measured in the optical,
while the mid-infrared spectrum
shows the spectral characteristics of an AGN (Taniguchi {\it{et al.}}~1999) and
the presence of the starburst-like 7.7 $\mu$m PAH (polycyclic aromatic 
hydrocarbon) emission (Genzel {\it{et al.}}~1998; Lutz, Veilleux, \& Genzel 1999).

Based on our new results and previously published results, we 
hypothesize that the optical LINER-like nucleus is in fact a region of the 
galaxy close to, and ionized by, the extended (starburst?) near-infrared nucleus while the
Seyfert 2 nebula is directly ionized by the $K$-band point-like nucleus 
(dust-enshrouded
AGN?).

\subsection{Detection and Mapping of High-Velocity Gas Flows}

The profiles of the [OIII] 5007\AA\ emission line show substructures like
double peaks, blue and red bumps, indicating the presence of at least one additional 
velocity component in a large region around the LINER-like nucleus (see Figure 1 to
visualize the changes in the profile). 
We have decomposed these profiles into two gaussian components in order to detect the presence 
of coherent gas flows over large regions and to map their spatial distribution 
and velocity field. The light distribution and velocity field of these two
components are presented in Figure 2 (middle and right panels, respectively),
where the [OIII] velocity 
 measured in the optical nucleus (11250 km s$^{-1}$) has been subtracted.

The primary [OIII] component (called component A in Figure 2) has a spatial distribution
 coincident with that of the 
low-excitation H$\beta$-emitting gas, including the low surface brightness 
tongue located south of the brightest [OIII]-emitting peak. The velocities displayed
by this gas have an amplitude (peak-to-peak) of about 400 km s$^{-1}$. 
This [OIII] component also has a wide profile with values (FWHM) ranging from 
$\sim$ 400 to 600 km s$^{-1}$, similar to the line widths of several other luminous and 
ultraluminous infrared galaxies 
(Heckman {\it{et al.}}~1990). 
Thus, the kinematics of the ionized gas associated with the optical 
LINER nucleus 
and surrounding regions is very complex and appears to be highly disturbed.   
 
The secondary [OIII] component (called component B in Figure 2) is energetically a 
minor contributor, but on the contrary shows 
large velocities (see bottom right panel in Figure 2) and has a spatial structure 
well differentiated from that of the [OIII] primary component and low-excitation gas
(see bottom middle panel in Figure 2). The velocity field of this component has one
of the largest velocity amplitudes (2.4 $\times 10^3$ km s$^{-1}$,  peak-to-peak) 
measured in luminous infrared galaxies (cf.~NGC 3079; Veilleux {\it{et al.}}~1994) and
ULIRGs. The maximum velocity gradient of the flow is detected along PA135, 
and its dynamical center does not coincide
with the optical nucleus, but it is offset by about 1$'' \pm 0.5''$ 
to the southwest. The extended
near-infrared nucleus observed in the NICMOS images (Scoville {\it{et al.}}~1999) is located
about 0.6$''$ southwest (PA225) of the optical nucleus. If this $K$-band nucleus traces a nuclear 
starburst, the high-velocity flows detected here could then be the associated 
starburst-driven superwind similar to those detected in other luminous infrared 
galaxies (Heckman {\it{et al.}}~1990; Colina, Lipari, \& Macchetto 1991).
   
Finally, the ionized gas in the off-nucleus Seyfert 2 nebula
is blueshifted by 100$-$200 km s$^{-1}$ with respect to the velocity of the optical
nucleus, shows no evidence for coherent flows (peak-to-peak velocity amplitude
of less than 100 km s$^{-1}$), and has a narrow profile (FWHM $\leq 260$ km s$^{-1}$).
These properties are consistent with interstellar gas in a quiescent state, located in the outer 
regions of the galaxy, and not affected by either starburst- or AGN-related gas flows.
These characteristics are reminiscent of the ionized gas detected in the extended, kiloparsec
size, ionization cones of some Seyfert 2 galaxies.

\section{SUMMARY}

The excitation conditions and kinematics of the emission line gas in the ultraluminous 
infrared galaxy Mrk 273 have been investigated on the basis of integral field optical
spectroscopy.
The new data have shown that the ionized gas in Mrk 273 has an extended structure
formed by two main emitting regions. The first region coincides with the 
optical nucleus and has the spectral characteristics of a low-luminosity LINER. The 
second region is located at a distance of 4$''$ (3.1 kpc) southwest of the nucleus,
and is classified as a Seyfert 2. 
 
The profile of the [OIII] line presents double peaks indicating the existence of at least 
two dynamical components. The primary, energetically dominant, component 
reveals highly disturbed gas around the LINER nucleus. 
A secondary high-velocity gas flow (2.4 $\times$ 10$^3$ km s$^{-1}$, peak-to-peak amplitude) 
is seen as evidence for a starburst-driven superwind generated in a dust-obscured nuclear 
starburst. 

The [OIII] ionized gas in the Seyfert 2 nebula has the properties of quiescent interstellar gas 
located in the outer regions of the galaxy and being 
directly ionized by a dust-enshrouded AGN, similar to the ionization cones of Seyfert 2 galaxies.

\acknowledgments 
Luis Colina thanks the Instituto de Astrof\'{\i}sica de Canarias for
its hospitality and financial support.
Kirk Borne thanks Raytheon for providing financial support
during his Sabbatical Leave. 
Support for this work was provided by NASA through 
grant number GO-06346.01-95A from the Space
Telescope Science Institute, which is operated by 
AURA, Inc., under NASA contract NAS5-26555.

\newpage

\vspace*{0.5cm}

\figcaption{The final calibrated [OIII]5007\AA\ profiles obtained
   with INTEGRAL are shown . The bundle of fibers cover a projected size of 16.5$'' \times 12.3''$ 
   on the sky, and the numbers in the plot indicate the number of the fiber at the entrance
   of the spectrograph ({\it{i.e.}}, pseudo-slit).
   The substructure and double peaks in several of the fibers (69, 80, 91, 92, 112, 133, etc.)
   are clearly visible.}

\figcaption{Images of the stellar component and ionized gas as traced by the blue continuum
   (bottom left panel),
H$\beta$ line emission (top left panel), and [OIII]5007\AA\ 
line emission (middle panels).
The velocity fields for the two [OIII]5007\AA\ components are also shown
   (right panels), clearly indicating the high-velocity gas flows detected in component B
   (see $\S$3 for details). 
   The scales of each of the intensity maps were selected independently and are in arbitrary 
   units.}

\figcaption{Plot showing the spectral regions 
of fibers 95 and 120 corresponding to
   the peak emission of the two main 
line-emitting regions: the northeastern nuclear H$\beta$
   and the southwestern off-nucleus [OIII] regions, respectively. The spectra
   clearly show the LINER- and Seyfert2-like spectral characteristics for the nucleus and
   off-nucleus regions, respectively. Note that the horizontal axis 
does not represent a continuous range in wavelength, but two subsets 
corresponding to the H$\beta$/[OIII] and H$\alpha$/[NII] + [SII] emission lines}

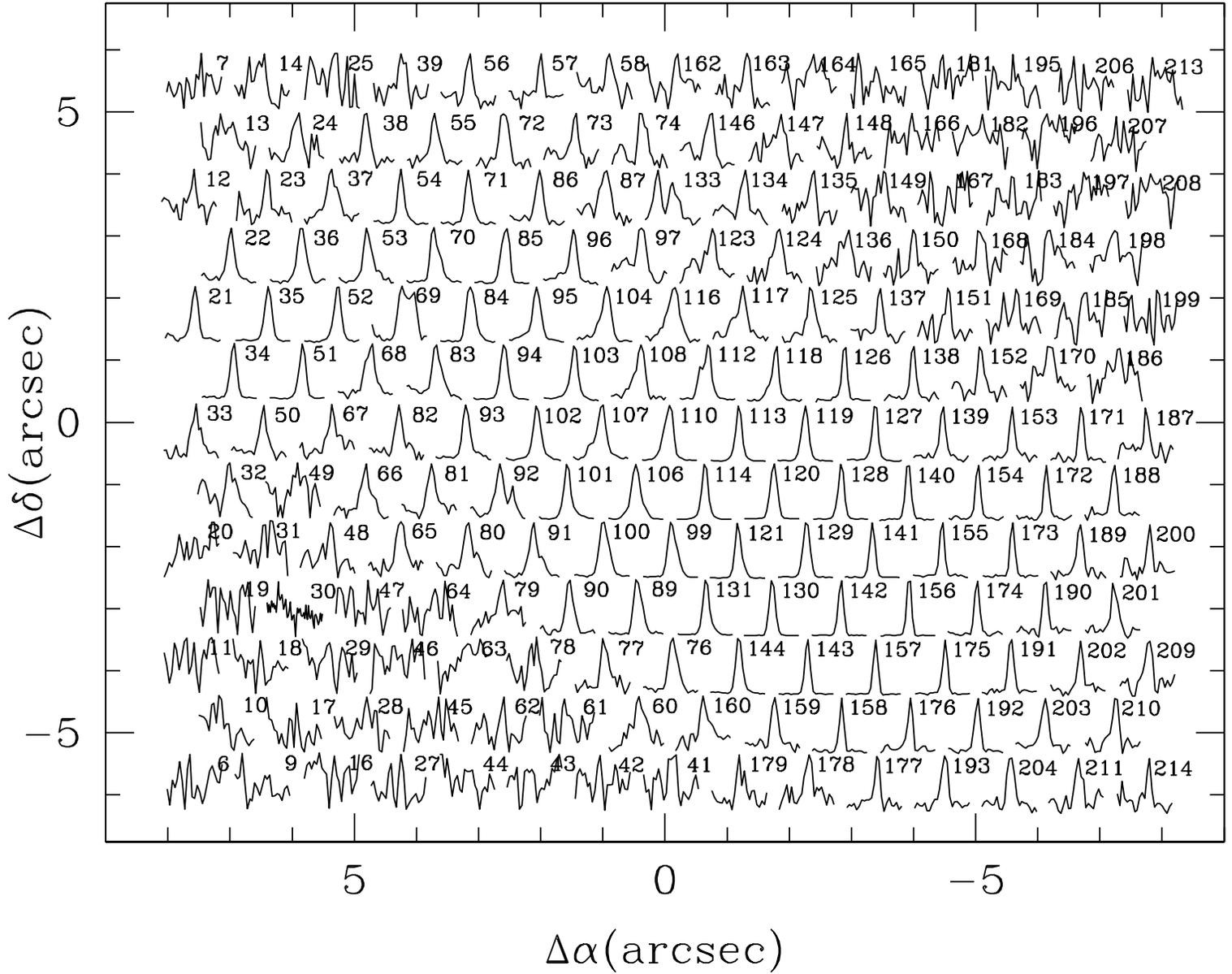
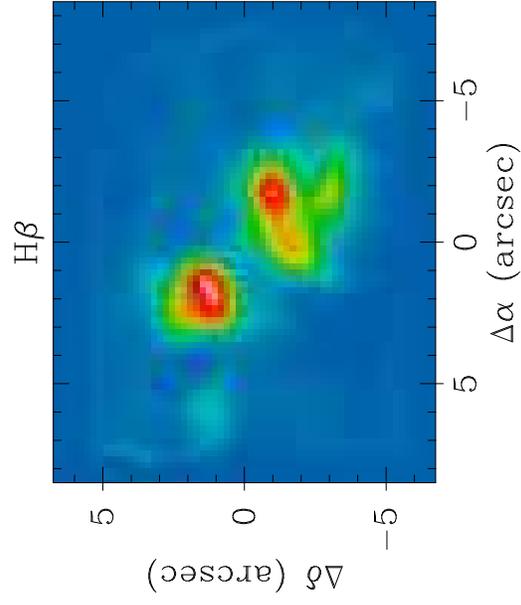
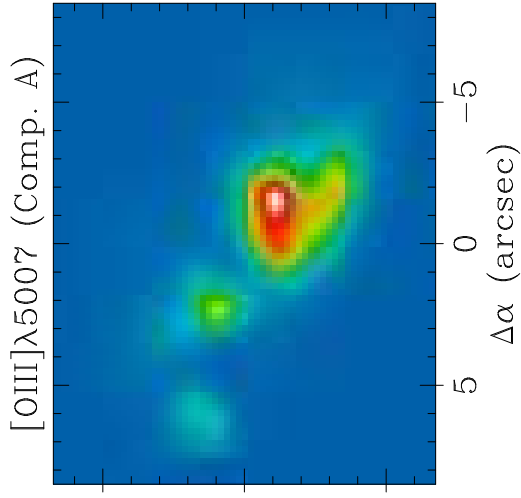
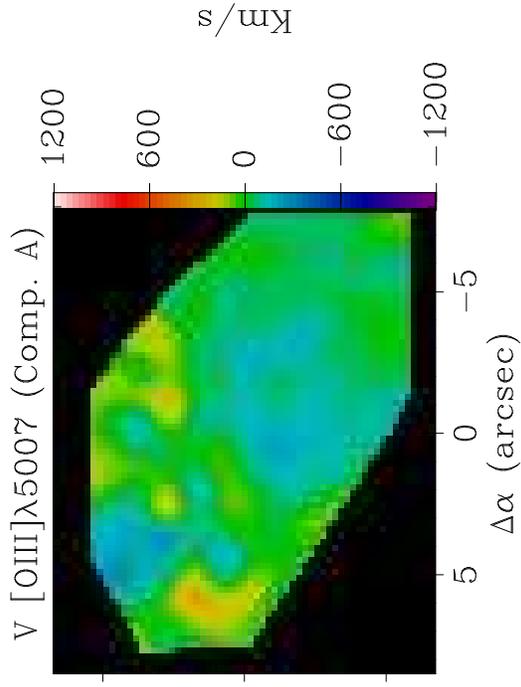
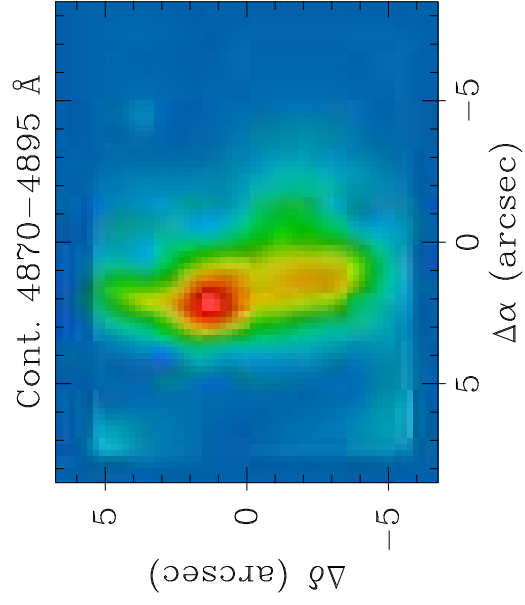
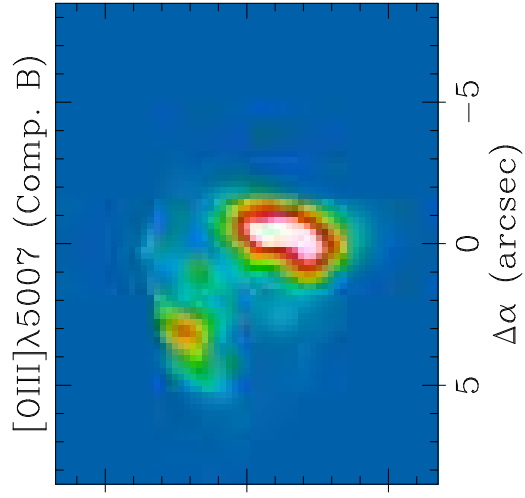
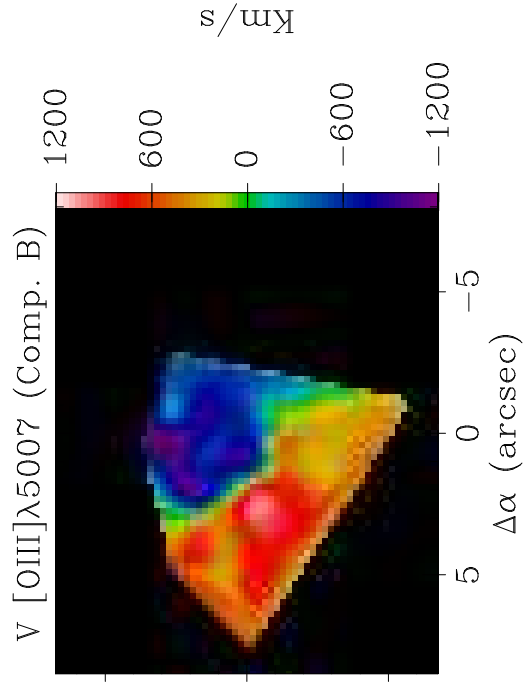
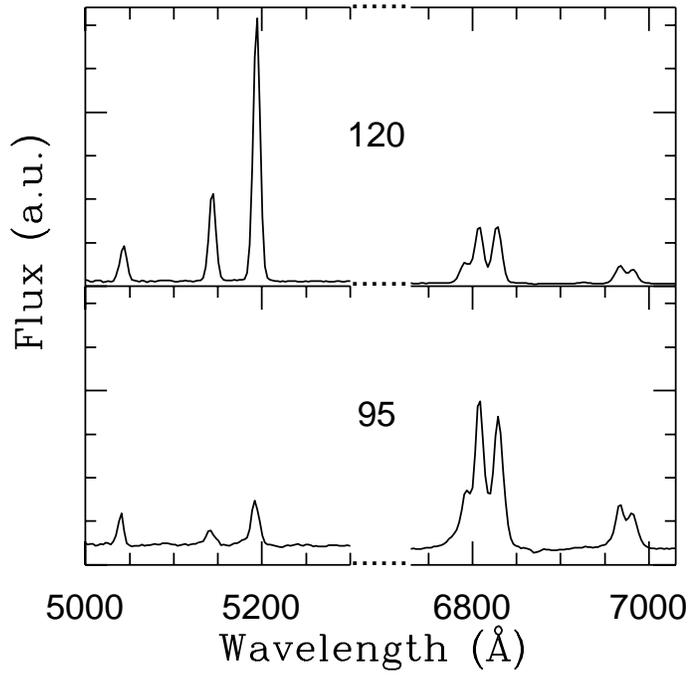
\end{document}